\documentclass[aps,prx,twocolumn,showpacs,superscriptaddress]{revtex4-1}
\usepackage{amsmath}
\usepackage{graphicx}
\usepackage{subfigure}
\usepackage{ulem,color}
\begin{document}
\title{Breaking the current density threshold in spin-orbit-torque
magnetic random access memory}
\author{Yin Zhang}
\affiliation{Physics Department, The Hong Kong University of Science and
Technology, Clear Water Bay, Kowloon, Hong Kong}
\affiliation{HKUST Shenzhen Research Institute, Shenzhen 518057, China}
\author{H. Y. Yuan}
\email{[Corresponding author:]yuanhy@sustc.edu.cn}
\affiliation{Department of Physics, Southern University of Science and
Technology of China, Shenzhen 518055, China}
\author{X. S. Wang}
\affiliation{School of Microelectronics and Solid-State Electronics,
University of Electronic Science and Technology of China, Chengdu,
Sichuan 610054, China}
\affiliation{Physics Department, The Hong Kong University of Science and
Technology, Clear Water Bay, Kowloon, Hong Kong}
\author{X. R. Wang}
\email{[Corresponding author:]phxwan@ust.hk}
\affiliation{Physics Department, The Hong Kong University of Science and
Technology, Clear Water Bay, Kowloon, Hong Kong}
\affiliation{HKUST Shenzhen Research Institute, Shenzhen 518057, China}
\date{\today}

\begin{abstract}
Spin-orbit-torque magnetic random access memory (SOT-MRAM) is a promising
technology for the next generation of data storage devices. The main
bottleneck of this technology is the high reversal current density threshold.
This outstanding problem of SOT-MRAM is now solved by using a current density of
constant magnitude and varying flow direction that reduces the reversal current
density threshold by a factor of more than the Gilbert damping coefficient.
The Euler-Lagrange equation for the fastest magnetization reversal path
and the optimal current pulse are derived for an arbitrary magnetic cell.
The theoretical limit of minimal reversal current density and current density
for a GHz switching rate of the new reversal strategy for CoFeB/Ta SOT-MRAMs
are respectively of the order of $10^5$ A/cm$^2$ and $10^6$ A/cm$^2$ far
below $10^7$ A/cm$^2$ and $10^8$ A/cm$^2$ in the conventional strategy.
Furthermore, no external magnetic field is needed for a deterministic reversal
in the new strategy.
\\

Subject Areas: Magnetism, Nanophysics, Spintronics

\end{abstract}

\maketitle

\section{Introduction}
Fast and efficient magnetization reversal is of not only fundamentally
interesting, but also technologically important for high density data
storage and massive information processing. Magnetization reversal can
be induced by magnetic field \cite{Back,Bauer,zzsun2006}, electric
current through direct \cite{Slonczewski,Berger,Katine,szhang2004,
Wetzels,zzsun2007} and/or indirect \cite{Hirsch,cornell,Beach,Lee,
Fukami,chen,szhang2009,Miron,xfhan,koopmans,Fukami2,Manchon,Miron2}
spin angular momentum transfer from polarized itinerant electrons to
magnetization, microwaves \cite{zzsun-mw}, laser light \cite{Bigot},
and even electric fields \cite{Ohno}. While the magnetic field
induced magnetization reversal is a matured technology, it suffers
from scalability and field localization problems \cite{Wetzels,stram}
for nanoscale devices. Spin transfer torque magnetic random-access
memory is an attractive technology in spintronics \cite{stram}
although Joule heating, device durability and reliability are
challenging issues \cite{stram,cornell}. In an spin-orbit-torque
magnetic random access memory (SOT-MRAM) whose central component is
a heavy-metal/ferromagnet bilayer, an electric current in the
heavy-metal layer generates a pure spin current through the spin-Hall
effect \cite{Hirsch,cornell} that flows perpendicularly into the
magnetic layer. The spin current, in turn, produces spin-orbit torques
(SOT) through spin angular momentum transfer \cite{Slonczewski,Berger}
and/or Rashba effect \cite{szhang2009,Miron,xfhan,koopmans,Fukami2,
Manchon,Miron2}. SOT-MRAM is a promising technology because writing
charge current does not pass through the memory cells so that the cells
do not suffer from the Joule heating and associated device damaging.
In principle, such devices are infinitely durable due to negligible
heating from spin current \cite{cornell}. However, the reversal
current density threshold (above $10^7$ A/cm$^2$ \cite{Fukami,chen} for
realistic materials) in the present SOT-MRAM architecture is too high.
To have a reasonable switching rate (order of GHz), the current
density should be much larger than $10^8$ A/cm$^2$ \cite{Fukami,chen}
that is too high for devices. In order to lower the minimal reversal
current density as well as to switch magnetization states at GHz rate
at a tolerable current density in SOT-MRAM, it is interesting to
find new reversal schemes (strategies) that can achieve above goals.
In this paper, we show that a proper current density pulse of
time-dependent flow direction and constant magnitude, much lower
than the conventional threshold, can switch a SOT-MRAM at GHz rate.
Such a time-dependent current pulse can be realized by using two
perpendicular currents passing through the heavy-metal layer.
The theoretical limit of minimal reversal current density of the
new reversal strategy for realistic materials can be of the order of
$10^5$ A/cm$^2$, far below $10^7$ A/cm$^2$ in the conventional
strategy that uses a direct current (DC), both based on macrospin
approximation. The validity of the macrospin model is also verified
by micromagnetic simulations.

\begin{figure}
\centering
\includegraphics[width=0.45\textwidth]{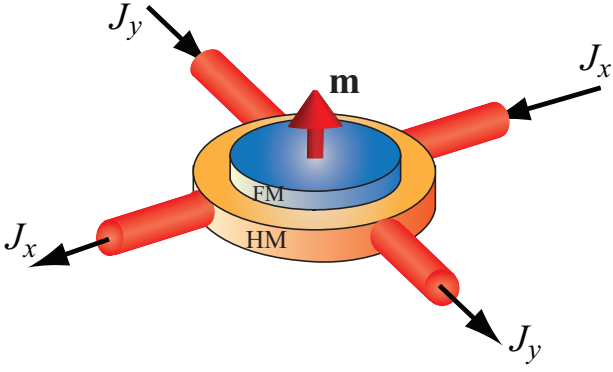}\\
\caption{Schematic illustration of new reversal scheme for SOT-MRAMs.
Two perpendicular currents flow in the heavy-metal layer of a
ferromagnet/heavy-metal bilayer to generate a current whose
direction can vary in the $xy$-plane.}
\label{configs}
\end{figure}

\section{Macrospin Model and Results}

\subsection{Model}
Our new reversal strategy for an SOT-MRAM, whose central component is a
ferromagnetic/heavy-metal bilayer lying in the $xy$-plane with initial
spin along the $+z$-direction as shown in Fig. \ref{configs}, uses a
current density $\mathbf J=J\cos\Phi\hat x+J\sin\Phi\hat y$ generated
from two time-dependent electric currents flowing along the $x$- and
the $y$-directions, where $\Phi$ is a time-dependent angle between
$\mathbf J$ and the $x$-axis and $J$ is a constant total current density.
The magnetic energy density is $\varepsilon=-K\cos^2\theta$ with $K$
being the anisotropy coefficient and $\theta$ being the polar angle of
the magnetization. In the absence of an electric current, the system has
two stable states $\mathbf m=+\hat z$ and $\mathbf m=-\hat z$ where
$\mathbf m$ is the unit direction of magnetization $\mathbf M=M\mathbf m$
of magnitude $M$. The electric current generates a transverse spin
current perpendicularly flowing into the ferromagnetic layer via the
spin-Hall effect \cite{Hirsch}, and then produces an effective SOT on
the magnetization \cite{Slonczewski,Berger,szhang2009}, i.e.
\begin{equation}
\vec\tau=-a\mathbf m\times(\mathbf m\times\hat s)
+\beta a\mathbf m\times\hat s,
\label{spin-torque}
\end{equation}
where the first term on the right-hand-side is the Slonczewski-like
torque while the second term is the field-like torque.
The spin-polarization direction is $\hat s=\hat J\times\hat z$ (for
other type of spin-Hall effect, see Note \cite{note1}) with $\hat J$
being the unit vector of current density.
$a=\frac{\hbar}{2ed}\theta_\mathrm{SH}J$ measures SOT where
$\hbar$, $e$, and $d$ are respectively the Plank constant, the electron
charge, and the sample thickness. $\theta_\mathrm{SH}$ is the spin Hall
angle which measures the conversion efficiency between the spin current
and charge current. $\beta$ measures the field-like torque and can be an
arbitrary real number since this torque may also be directly generated
from the Rashba effect \cite{szhang2009}.

The magnetization dynamics under an in-plane current density $\mathbf J$
is governed by the generalized dimensionless Landau-Lifshitz-Gilbert
(LLG) equation,
\begin{equation}
\frac{\partial\mathbf m}{\partial t}=-\mathbf m\times
\mathbf h_\mathrm{eff} +\alpha\mathbf m
\times\frac{\partial\mathbf m}{\partial t}+\vec\tau,
\label{llg-eq}
\end{equation}
where $\alpha$ is the Gilbert damping constant that is
typically much smaller than unity. The effective field is
$\mathbf h_\mathrm{eff}=-\nabla_\mathbf m \varepsilon$
from energy density $\varepsilon$. Time, magnetic field and
energy density are respectively in units of $(\gamma M)^{-1}$,
$M$ and $\mu_0 M^2$, where $\gamma$ and $\mu_0$ are respectively
the gyromagnetic ratio and vacuum magnetic permeability.
In this unit system, $a=\frac{\hbar}{2ed\mu_0M^2}\theta_
\mathrm{SH}J$ becomes dimensionless.

The magnetization $\mathbf m$ can be conveniently described by
a polar angle $\theta$ and an azimuthal angle $\phi$ in the
$xyz$-coordinate.
In terms of $\theta$ and $\phi$, the generalized LLG equation becomes
\begin{widetext}
\begin{subequations}\label{llg-eq2}
\begin{align}
(1+\alpha^2)\dot\theta=
-\alpha K\sin2\theta+a(1-\alpha\beta)\cos\theta\sin(\Phi-\phi)
+a(\alpha+\beta)\cos(\Phi-\phi)\equiv F_1, \label{llg-eq2a}\\
(1+\alpha^2)\dot\phi\sin\theta=K\sin2\theta-a(1-\alpha\beta)
\cos(\Phi-\phi)+a(\alpha+\beta)\cos\theta\sin(\Phi-\phi)
\equiv F_2. \label{llg-eq2b}
\end{align}
\end{subequations}
\end{widetext}

\subsection{Derivation of the Euler-Lagrange equation}
The goal is to reverse the initial state $\theta=0$ to the target
state $\theta=\pi$ by SOT.
There are an infinite number of paths that connect the initial
state $\theta=0$ with the target state $\theta=\pi$, and each of
these paths can be used as a magnetization reversal route.
For a given reversal route, there are an infinite number
of current pulses that can reverse the magnetization.
The theoretical limit of minimal current density $J_c$ is
defined as the smallest values of minimal reversal current
densities of all possible reversal routes.
Then it comes two interesting and important questions: 1)
What is $J_c$ above which there is at least one reversal route
that the current density can reverse the magnetization along it?
2) For a given $J>J_c$, what are the optimal reversal route and
the optimal current pulse $\Phi(t)$ that can reverse the
magnetization at the highest speed?

Dividing Eq. \eqref{llg-eq2b} by Eq. \eqref{llg-eq2a}, one can
obtain the following constraint,
\begin{equation}
G\equiv \frac{\partial\phi}{\partial\theta}\sin\theta F_1-F_2 =0.
\label{constraint}
\end{equation}
The magnetization reversal time $T$ is
\begin{equation}
T=\int_0^\pi\frac{d\theta}{\dot\theta}=\int_0^\pi\frac{1+\alpha^2}{F_1}d\theta.
\label{reversal-time}
\end{equation}
The optimization problem here is to find the optimal reversal
route $\phi(\theta)$ and the optimal current pulse $\Phi(t)$
such that $T$ is minimum under constraint \eqref{constraint}.
Using the Lagrange multiplier method, the optimal reversal route
and the optimal current pulse satisfy the Euler-Lagrange
equations \cite{Wang2008,Arfken},
\begin{equation}
\begin{gathered}
\frac{\partial F}{\partial\phi}=\frac{d}{d\theta}(\frac{\partial
F}{\partial(\partial\phi/\partial\theta)}),
\frac{\partial F}{\partial\Phi}=\frac{d}{d\theta}(\frac{\partial
F}{\partial(\partial\Phi/\partial\theta)}),
\end{gathered}
\label{Euler-eqs}
\end{equation}
where $F=(1+\alpha^2)/F_1+\lambda G$ and $\lambda$ is the
Lagrange multipliers which can be determined self-consistently
by Eq. \eqref{Euler-eqs} and constrain \eqref{constraint}.
Given a current density of constant magnitude $J$, Eq.
\eqref{Euler-eqs} may or may not have a solution of $\phi(\theta)$
that continuously passing through $\theta=0$ and $\theta=\pi$.
If such a solution exists, then $\phi(\theta)$ is the optimal
path for the fastest magnetization reversal and the
corresponding solution of $\Phi(t)$ is the optimal current pulse.
The theoretical limit of minimal reversal current density is
then the smallest current density $J_c$ below which the optimal
reversal path does not exist.

\subsection{The optimal current pulse and theoretical limit of
minimal reversal current density}
From Eqs. \eqref{llg-eq2a}, \eqref{llg-eq2b} and \eqref{constraint}
as well as $F=(1+\alpha^2)/F_1+\lambda G$, the Euler-Lagrange equation
of \eqref{Euler-eqs} becomes
\begin{subequations}
\begin{align}
&\lambda\frac{d}{d\theta}(F_1\sin\theta)=0, \label{Euler-eqs2a}\\
&\frac{1+\alpha^2}{F_1^2}\frac{\partial F_1}{\partial\phi}
-\lambda\frac{\partial G}{\partial\phi}=-\frac{1+\alpha^2}{F_1^2}
\frac{\partial F_1}{\partial\Phi}
+\lambda\frac{\partial G}{\partial\Phi}=0. \label{Euler-eqs2b}
\end{align}
\end{subequations}
From Eq. \eqref{Euler-eqs2a}, one has $\lambda\neq 0$ or $\lambda=0$.
If $\lambda \neq 0$, $F_1$ must be $F_1=C/\sin\theta$ ($C\neq0$) so
that $(1+\alpha^2)\dot\theta=C/\sin\theta \rightarrow \infty$ as
$\theta\rightarrow 0$ or $\pi$. This solution is not physical, and
should be discarded. Therefore, the only allowed solution must be
$\lambda=0$, and one has $\partial F_1/\partial\Phi=0$ according to
Eq. \eqref{Euler-eqs2b}. Interestingly, this is exactly the condition
of maximal $\dot\theta=F_1/(1+\alpha^2)$ as $\Phi$ varies. $\Phi$ satisfies
$\tan(\Phi-\phi)=\frac{1-\alpha\beta}{\alpha+\beta}\cos\theta$, or
\begin{subequations}\label{optimal2}
\begin{align}
\Phi &= \tan^{-1}(\frac{1-\alpha\beta}{\alpha+\beta}\cos\theta)
+\phi+\pi \quad &(\beta<-\alpha)\\
\Phi &= \tan^{-1}(\frac{1-\alpha\beta}{\alpha+\beta}\cos\theta)
+\phi \quad &(\beta>-\alpha).
\end{align}
\end{subequations}

Substituting Eq. \eqref{optimal2} into the LLG equation \eqref{llg-eq2},
$\theta(t)$ and $\phi(t)$ are determined by the following equations,
\begin{subequations}
\begin{align}
\dot\theta &= \frac{1}{1+\alpha^2}[a P(\theta)-\alpha K
\sin 2\theta], \label{dot_theta}\\
\dot\phi &= \frac{1}{1+\alpha^2}[2K\cos\theta-a(\alpha+\beta)
(1-\alpha\beta)\frac{\sin\theta}{P(\theta)}], \label{dot_phi}
\end{align}
\end{subequations}
where $P(\theta)=\sqrt{(\alpha+\beta)^2+(1-\alpha\beta)^2\cos^2\theta}$.
To reverse magnetization from
$\theta=0$ to $\theta=\pi$, $a$ must satisfy $a >\alpha K \sin
(2\theta)/P(\theta)$ according to Eq. \eqref{dot_theta}
so that $\dot{\theta}$ is no negative for all $\theta$.
Obviously, $\dot\theta=0$ at $\theta=\pi/2$ when $\beta=-\alpha$.
The magnetization reversal is not possible in this case, and
$\beta=-\alpha$ is a singular point. The theoretical limit of minimal
reversal current density $J_c$ for $\beta\neq -\alpha$ is
\begin{equation}
J_c=\frac{2\alpha eKd}{\theta_\mathrm{SH}\hbar}Q,
\label{J_c}
\end{equation}
where $Q\equiv\mathrm{max}\{\sin 2\theta/P(\theta)\}$ for $\theta\in[0,\pi]$.

\begin{figure}
\centering
\includegraphics[width=0.4\textwidth]{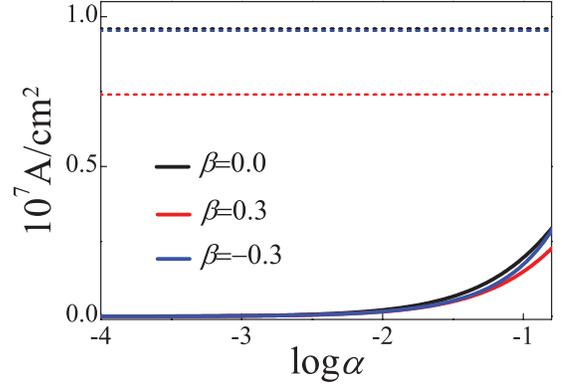}\\
\caption{
The $\log\alpha$-dependence of $J_c$ for various $\beta$ are plotted
as the solid curves for model parameters of $M=3.7\times10^5$ A/m,
$K=5.0\times10^3$ J/m$^3$, $\theta_\mathrm{SH}=0.084$ and $d=0.6$ nm.
As a comparison, $J_c^\mathrm{dc}$ is also plotted as the dashed lines.}
\label{fig_J_c}
\end{figure}

In comparison with the current density threshold \cite{Lee,Fukami,xfhan}
($J_c^\mathrm{dc}$) in the conventional strategy for $\beta=0$,
\begin{equation}
J_c^\mathrm{dc}=\frac{2eKd}{\theta_\mathrm{SH}\hbar}
(1-\frac{H}{\sqrt{2}K}),
\label{J_c_dc}
\end{equation}
the minimal reversal current density is reduced by more than a factor
of $\alpha$. Here $H\ (\simeq 22$ Oe in experiments) is a small external
magnetic needed for a deterministic reversal in conventional strategy.
Using CoFeB/Ta parameters of $M=3.7\times10^5$ A/m, $K=5.0\times10^3$ J/m$^3$,
$\theta_\mathrm{SH}=0.084$ and $d=0.6$ nm \cite{cornell,Fukami,chen},
Fig. \ref{fig_J_c} shows $\log\alpha$-dependence of $J_c$ (solid
lines) and $J_c^\mathrm{dc}$ (dashed lines) for $\beta=0$ (black),
$0.3$ (red) and $-0.3$ (blue), respectively.
Both $J_c^\mathrm{dc}$ and $J_c$ depend on $\beta$.
The lower the damping of a magnetic material is, the smaller our
minimum switching current density will be. For a magnetic material
of $\alpha=10^{-5}$, the theoretical limit of minimal reversal
current density can be five order of magnitude smaller than the value
in the conventional strategy.

\begin{figure*}
\centering
\includegraphics[width=0.9\textwidth]{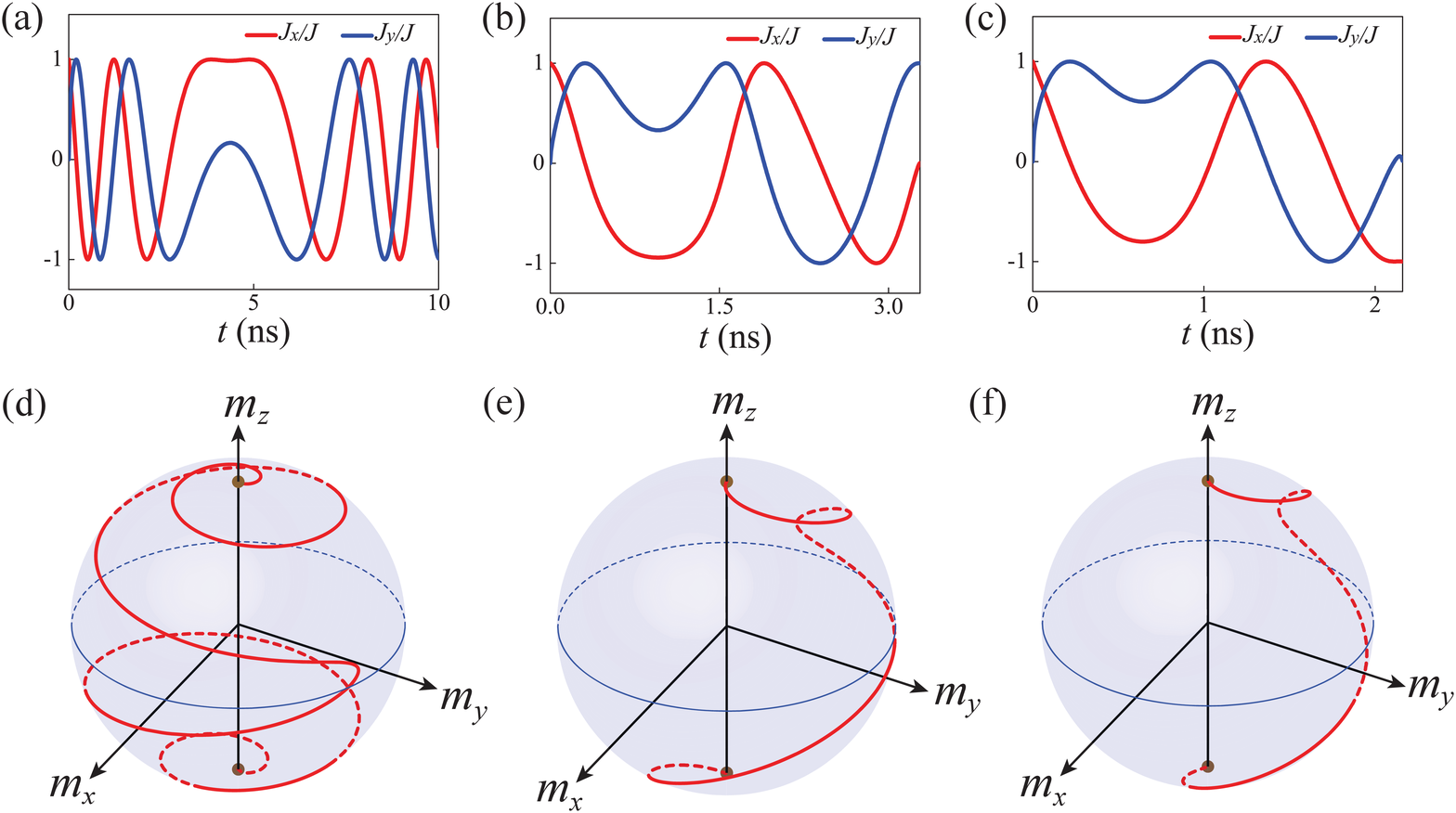}\\
\caption{Model parameters of $M=3.7\times10^5$ A/m,
$K=5.0\times10^3$ J/m$^3$, $\theta_\mathrm{SH}=0.084$, $\alpha=0.008$
and $d=0.6$ nm are used to mimic CoFeB/Ta bilayer, and
$\beta=0.3$ for (a), (c), (d) and (f) while $\beta=0.1$ for (b) and (e).
The theoretical limit of minimum reversal current density is
$J_c=1.56\times10^5$ A/cm$^2$ for $\beta=0.1$ and
$J_c=1.28\times10^5$ A/cm$^2$ for $\beta=0.3$. Optimal current pulses
((a)-(c)) and fastest reversal routes ((d)-(f)) are
for $J=1.92\times10^6$ A/cm$^2$ ((a) and (d)), and
for $J=9.0\times10^6$ A/cm$^2$ ((b), (c)), (e) and (f).}
\label{fig_pulse}
\end{figure*}

\begin{figure}
\centering
\includegraphics[width=0.4\textwidth]{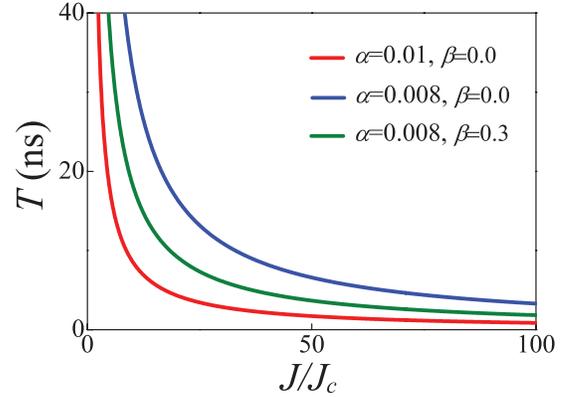}\\
\caption{Magnetization reversal time $T$ under the optimal current pulses
as a function of $J$ for various $\alpha$ and $\beta$.}
\label{fig_time}
\end{figure}

For a given $J>J_c$, the shortest reversal time is given by Eqs.
\eqref{reversal-time} and \eqref{dot_theta}:
\begin{equation}
T=\int_0^\pi \frac{1+\alpha^2}{aP(\theta)-\alpha K\sin2\theta} d\theta.
\end{equation}
The optimal reversal path is given by $\phi(\theta)=\int_0^
\theta\frac{\dot\phi}{\dot\theta}d\theta'$ where $\dot\theta$ and
$\dot\phi$ are given by Eqs. \eqref{dot_theta} and \eqref{dot_phi}.
Eq. \eqref{dot_theta} gives $t(\theta)=\int_0^
\theta(1+\alpha^2)/(aP(\theta)-\alpha K\sin2\theta) d\theta'$
and then $\theta(t)$ is just $\theta(t)=t^{-1}(\theta)$.
Thus, $\Phi(\theta,\phi)$, $\phi(\theta)$ and $\theta(t)$ give
$\phi(t)=\phi(\theta(t))$ and $\Phi(t)=\Phi(\theta(t),\phi(t))$.
Using the same parameters as those for Fig. \ref{fig_J_c} with
$\alpha=0.008$ and various $\beta$, Fig. \ref{fig_pulse} shows the optimal
current pulses ((a)-(c)) and the corresponding fastest magnetization
reversal routes ((d)-(f)) for $\beta=0.3$ and $J=1.92\times10^6$ A/cm$^2\approx 15 J_c$
((a) and (d)), for $\beta=0.1$ and $J=9.0\times10^6$ A/cm$^2\approx 58 J_c$
((b) and (e)), and for $\beta=0.3$ and $J=9.0\times10^6$ A/cm$^2\approx 70 J_c$
((c) and (f)). It is known that Ta has less effect on $\alpha$ \cite{cornell}.
The minimal reversal current density $J_c$ under the optimal current
pulse is $1.56\times10^5$ A/cm$^2$ for $\beta=0.1$ and $1.28\times10^5$ A/cm$^2$
for $\beta=0.3$ which is far below $J_c^\mathrm
{dc}=9.6\times10^6$ A/cm$^2$ for the same material parameters \cite{chen}.
The multiple oscillations of $m_x$ and $m_y$ reveal that the reversal is
a spinning process and optimal reversal path winds around the two stable
states many times. Correspondingly, the driving current makes also
many turns as shown by the multiple oscillations of $J_x$ and $J_y$.
The number of spinning turns depends on how far $J$ is from $J_c$.
The closer $J$ to $J_c$ is, the number of turns is larger. The number
of turns is about 5 in Figs. \ref{fig_pulse}(a) and \ref{fig_pulse}(d) for
$J\approx 15J_c$ and one turn for $J>50J_c$ as shown in
Figs. \ref{fig_pulse}(b), \ref{fig_pulse}(c), \ref{fig_pulse}(e) and
\ref{fig_pulse}(f), so that the reversal is almost ballistic.
The reversal time for $\beta=0.3$ and $J=1.92\times10^6$ A/cm$^2$
is about 10 nanoseconds, for $\beta=0.1$ and $J=9.0\times10^6$ A/cm$^2$
is about 3.3 nanoseconds, and for $\beta=0.3$ and $J=9.0\times10^6$ A/cm$^2$
is about 2.1 nanoseconds.
Figure \ref{fig_time} is the reversal time $T$ as a
function of current density $J$ under the optimal current pulse for
the same parameters as those for Fig. \ref{fig_J_c}.
The reversal time quickly decreases to nanoseconds as current density
increases. In a real experiment, there are many uncertainties
so that the current pulse may be different from the optimal one.
To check whether our strategy is robust again small fluctuations,
we let the current pulse in Fig. \ref{fig_pulse}(c) deviate from its
exact value. Numerical simulations show that the magnetization reversal
is not significantly influenced at least when the deviation between the
real current and optimal current is less than five percents.

\section{Verification of macrospin model by micromagnetic simulation}

\begin{figure}
\centering
\includegraphics[width=0.5\textwidth]{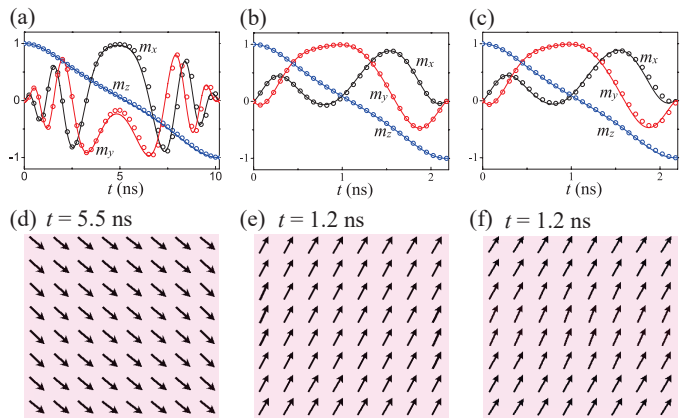}\\
\caption{(a)-(c) Time evolution of the average magnetization:
cycles for micromagnetic simulations and solid lines are theoretical
predictions from macrospin model. (a) and (b) are for the memory cell of
150 nm$\times$150 nm$\times$0.6 nm and optimal current pulse of current
density of $J=1.92\times10^6$ A/cm$^2$ and $J=9.0\times10^6$ A/cm$^2$,
respectively. (c) is  for the memory cell of 250 nm$\times$250 nm$
\times$0.6 nm and optimal current pulse of current
density of $J=9.0\times10^6$ A/cm$^2$.
(d)-(f) Spin configurations respectively corresponding to (a)-(c) in
the middle of magnetization reversal at $t=5.5$ ns and 1.2 ns.
The cell size in micromagnetic simulation is 2 nm$\times$2 nm$\times$0.6 nm.
}
\label{micromagnetic}
\end{figure}

In our analysis, the memory cell is treated as a macrospin. A nature
question is how good the macrospin model is for a realistic memory device.
To answer this question, we carried out micromagnetic simulations
by using Newton-Raphson algorithm \cite{daquino} for two memory cells of
150 nm$\times$150 nm$\times$0.6 nm (Figs. \ref{micromagnetic}(a), (b),
(d) and (e)) and 250 nm$\times$250 nm$\times$0.6 nm (Figs.
\ref{micromagnetic}(c) and (f)). To model the possible edge pinning effect
due to magnetic dipole-dipole interaction, we consider square-shape devices
instead of cylinder shape device whose edge pinning is negligible.
To make a quantitative comparison, the material parameters are the same
as those used in Fig. \ref{fig_pulse}. In our simulations, the unit cell
size is 2 nm$\times$2 nm$\times$0.6 nm. For a fair comparison, the optimal
current pulses shown in Figs. \ref{fig_pulse}(a) and (c) of respective
current density $J=1.92\times10^6$ A/cm$^2$ and $J=9.0\times10^6$ A/cm$^2$
were applied to the memory cell of 150 nm$\times$150 nm$\times$0.6 nm.
The symbols in Figs. \ref{micromagnetic}(a) and (b) are the time evolution
of averaged magnetization $m_x$, $m_y$ and $m_z$ while the solid lines are
the theoretical predictions of macrospin model shown in Figs.
\ref{fig_pulse}(d) and (f). The perfect agreements prove the validity of the
macrospin approximation for our device of such a size. To further verify that
the memory device can be treated as a macrospin, Figs. \ref{micromagnetic}(d)
and (e) are the spin configurations in the middle of the reversal at $t=5.5$ ns
for Fig. \ref{micromagnetic}(a) and at $t=1.2$ ns for Fig. \ref{micromagnetic}(b).
The fact that all spins align almost in the same direction verifies
the validity of the macro spin model. In real experiments, non-uniformity of
current density is inevitable. To demonstrate the macrospin model is
still valid, we let current density linearly varies from
$9.5\times10^6$ A/cm$^2$ on the leftmost column of cells to
$8.5\times10^6$ A/cm$^2$ on the rightmost column of cells.
As expected, there is no noticeable difference with the data shown in
Figs. \ref{micromagnetic}(b) and (e).

For the large memory device of 250 nm$\times$250 nm$\times$0.6 nm, the
optimal current pulse shown in Fig. \ref{fig_pulse}(c) of current density
$J=9.0\times10^6$ A/cm$^2$ was considered. The time evolution of averaged
magnetization $m_x$, $m_y$ and $m_z$ are plotted in Fig. \ref{micromagnetic}(c),
with the symbols for simulations and solid lines for the macrospin model.
They agree very well although there is a small deviation for device of
such a large size. Figure \ref{micromagnetic}(f) is the spin configurations
in the middle of the reversal at $t=1.2$ ns for Fig. \ref{micromagnetic}(c).
The marcospin model is not too bad although all spins are not perfectly aligned
in this case.

In summary, for a normal SOT-MRAM device of size less than 300 nm
\cite{cornell, chen}, macrospin model describes magnetization reversal
well. However, for a larger sample size and lower current density
($J<10^6$ A/cm$^2$ for the same material parameters as those used in
Fig. \ref{fig_pulse}), only the spins in sample center can be reversed
while the spins near sample edges are pinned.

\section{Discussion}

Obviously, the strategy present here can easily be generalized to
the existing spin-transfer torque MRAM. The mathematics involved
are very similar, and one expects a substantial current density
reduction is possible there if a proper optimal current pulse is used.
Of course, how to generate such a current pulse should be much more
challenge than that for SOT-MRAM where two perpendicular currents can be used.
In the conventional strategy that uses a DC-current, a static
magnetic field along current flow is required for a deterministic
magnetization reversal \cite{Lee,Fukami,xfhan}.
Although several field-free designs have been proposed
\cite{koopmans,Fukami2}, an antiferromagnet is needed to create an
exchange bias which plays the role of an applied magnetic field.
As we have shown, such a requirement or complication is not needed
in our strategy. Our strategy does not have another problem existing
in the conventional strategy in which the magnetization can only be
placed near $\theta=\pi/2$ \cite{Lee, Fukami, xfhan} so that the
system falls into the target state by itself through the damping.
Therefore, one would like to use materials with larger damping in
the conventional strategy in order to speed up this falling process.
In contrast, our strategy prefers low damping materials, and reversal
is almost ballistic when current density is large enough ($>50J_c$ in
the current case).
To reverse the magnetization from $\theta=\pi$ to $\theta=0$, one only
needs to reverse the current direction of the optimal current pulse.
One should notice that the Euler-Lagrange equation allows us to easily
obtain the optimal reversal current pulse and theoretical limit of the
minimal reversal current density for an arbitrary magnetic cell such
as in-plane magnetized layer \cite{cornell} and biaxial anisotropy.

\section{Conclusion}
In conclusion, we investigated the magnetization reversal of
SOT-MRAMs, and propose a new reversal strategy whose minimal reversal
current density is far below the existing current density threshold.
For popular CoFeB/Ta system, it is possible to use a current density
less than $10^6$ A/cm$^2$ to reverse the magnetization at GHz rate,
in comparison with order of $J\approx 10^8$ A/cm$^2$ in the
conventional strategy.

\begin{acknowledgments}
This work was supported by the National Natural Science Foundation
of China (Grant No. 11774296 and No. 61704071) as well as Hong Kong
RGC Grants No. 16300117 and No. 16301816. X.R.W. acknowledges the
hospitalities of Beijing Normal University and Beijing Computational
Science Research Center during his visits.
\end{acknowledgments}

\end{document}